\documentclass[reprint,%
 amssymb, amsmath, pre,
 aps, floatfix,
]{revtex4-1}

\usepackage{bm}%
\usepackage[colorlinks=true,linkcolor=blue]{hyperref}%
\usepackage{xcolor}

\usepackage{dcolumn}%
\usepackage{graphicx}
\expandafter\ifx\csname package@font\endcsname\relax\else
 \expandafter\expandafter
 \expandafter\usepackage
 \expandafter\expandafter
 \expandafter{\csname package@font\endcsname}%
\fi
\usepackage{subfigure}
\begin{document}

\title{Competition between entropy and energy in network glass: the hidden connection between intermediate phase and liquid-liquid transition }%
 
\author{J. Quetzalc\'oatl Toledo-Mar\'in}%
\email{j.toledo.mx@gmail.com}
\affiliation{Departamento de Sistemas Complejos, Instituto de
F\'{i}sica, Universidad Nacional Aut\'{o}noma de M\'{e}xico (UNAM),
Apartado Postal 20-364, 01000 M\'{e}xico, CDMX,
M\'{e}xico}%

\author{Le Yan}%
\email{lyan@kitp.ucsb.edu}
\affiliation{Kavli Institute for Theoretical Physics, University of California, Santa Barbara, CA 93106, USA}%

\date{\today}

\begin{abstract}
In network glass including chalcogenides, the network topology of microscopic structures can be tuned by changing the chemical compositions. As the composition is varied, an intermediate phase (IP) singularly different from the adjacent floppy or rigid phases on sides has been revealed in the vicinity of the rigidity onset of the network. 
Glass formers in the IP appear to be reversible at glass transition and strong in dynamical fragility. 
Meanwhile, the calorimetry experiments indicate the existence of a first-order liquid-liquid transition (LLT) at a temperature above the glass transition in some strong glass-forming liquids. 
How are the intermediate phase and the liquid-liquid transition related? 
Recent molecular dynamic simulations hint that the intermediate phase is thermodynamically distinct that the transitions to IP as varying the chemical composition in fact reflect the LLT: 
out of IP, the glass is frozen in vibrational entropy-dominated heterogeneous structures with voids; while inside IP, energy dominates and the microscopic structures of liquids become homogeneous. 
Here we demonstrate such first-order thermodynamic liquid-liquid transition numerically and analytically in an elastic network model of network glass and discuss possible experimental approaches to testify the connection. 
\end{abstract}

\pacs{}

\maketitle

\section*{Introduction}  

%
In network glass, the material properties relying on structures can be tuned by changing the chemical compositions that have different abilities to make covalent connections with its neighbor atoms. 
In chalcogenides Ge$_x$As$_y$Se$_{1-x-y}$, for example, selenium (Se) forms only two bonds while arsenic (As) and germanium (Ge) form three and four respectively. 
First pointed out by Maxwell~\cite{Maxwell64}, a general network will lose rigidity as the network connectivity is reduced to below certain critical connectivity when the average number of constraints per atom, $n$, is equal to the degrees of freedom, i.e., $n_c=d$ in spatial dimension $d$. 
This rigidity loss also applies to the chalcogenides when selenium concentration is high, predicted by Phillips in~\cite{Phillips79,Thorpe85}, where he showed that counting both radial and angular constraints of covalent bonds gives $n_{\rm Se}=2$, $n_{\rm As}=9/2$, and $n_{\rm Ge}=7$, indicating a chalcogenide glass is marginally rigid at a composition with average number of covalent bonds $r_c=2.4$. 
Since then, more and more works have shown that the thermodynamic and dynamic features of glass-forming liquids (not limited to chalcogenides) are strongly regulated by the rigidity transition of the microscopic networks~\cite{Hall03, Shintani08, Mauro09,Yan13}. 
One of the most interesting discoveries is the intermediate phase (IP) near $r_c$~\cite{Boolchand01,Wang05,Rompicharla08,Bhosle12}, which remains a big puzzle.

The intermediate phase appears to be singularly distinct from the adjacent rigid or floppy phases: the non-reversible heat, a glass-transition equivalent of the latent heat, vanishes~\cite{Boolchand01}; the stress heterogeneity disappears~\cite{Wang05,Rompicharla08};  the molar volume and fragility are sharply smaller~\cite{Bhosle12}. All available pieces of evidence suggest that the glass undergoes some transitions  when entering IP from either side~\cite{Bhosle12}. 
However, both, Maxwell's rigidity theorem and the rigidity percolation theory that takes into account fluctuations of random networks~\cite{Jacobs95,Jacobs96,Barre05} predict only a single transition in network constraint number $n$. 
Noticed the interval of the two rigidity transition points in two theories, Thorpe and his colleagues proposed a self-organized transition scenario, which predicts a rigidity window in between two transitions -- one corresponding to the loss of percolating rigidity as in the rigidity percolation theory and the other corresponding to the loss of ability to relax stress as in the Maxwell's theorem~\cite{Thorpe00,Chubynsky06}. 
This stress-free rigidity window relies on a subtle balance between the fluctuation or entropy facilitating the rigidity percolation and the energy eliminating the stress throughout the range, 
which is, however, fragile to the ubiquitous perturbations such as Van der Waals (VdW) forces and temperature~\cite{Yan14}. 
 {Despite in a more recent paper~\cite{Kirchner19}, Kirchner and Mauro provide a robust approach of computing the constraint number to determine IP in the presence of finite temperature, }  
the heterogeneous nature captured by a diverging correlation length at $n=n_c$~\cite{Briere07}, in fact, still contradicts the observations of a homogeneous IP. 

An alternative set of theoretical insights on IP is from the molecular dynamics simulations~\cite{Micoulaut13,Bauchy14,Bauchy15}, where a similar intermediate range of homogeneous structures is revealed by continuously tuning the pressure instead of composition. 
In the simulation, as the pressure gradually increases, the amorphous structure undergoes a liquid-liquid transition (LLT) from a more structured low-modulus low-density amorphous phase to a more homogeneous rigid high-density phase~\cite{Yildirim18}. When the composition is varied, the transition pressure shows a non-monotonic pattern with a lower value in an intermediate range near $r_c$, same as the pattern of the stress percolation pressure in chalcogenides~\cite{Wang05}. 
In addition, in experiments, a transition between two thermodynamically different liquids is indicated by a lambda peak in specific heat at a temperature above the glass transition in some strong glass-formers close to the rigidity threshold, including silica (${\rm SiO}_2$)~\cite{Angell08, Wei11, Angell11, Wei13}. 
These materials imply that the glass in IP may be rather in a different thermodynamic phase resulted from a transition above the glass transition and the transitions to IP directly reflect such liquid-liquid transitions. 
So what are the two different liquid phases in network glass? 

In the previous work~\cite{Yan18}, one of the authors showed with a network model that the vibrational entropy facilitates the rigid-floppy separated heterogeneous network structures close to the rigidity transition $n_c$ as floppy modes store large amounts of vibrational entropy~\cite{Naumis05} while cost little configurational entropy in marginally rigid networks. On the opposite, the elastic energy is lower in homogeneous structures with stresses evenly distributed~\cite{Yan14}. So under cooling, a network near $n_c$ inevitably undergoes a first-order transition from an entropy-dominated heterogeneous phase to an energy-dominated homogeneous phase. 
The interplay of the glass transition temperature $T_g$ and the LLT temperature $T_{ LLT}$ would then be key in determining which liquid phase the material is frozen in at the glass transition and all the consequential features. 
Here, we investigate the transition separating the two liquid phases by studying the thermodynamics of the same network model. 
We show that the network undergoes a first-order phase transition where free energy crosses over, associated energy and entropy are discontinuous, and specific heat jumps in the thermodynamic limit. 
Finally, we discuss how this liquid-liquid transition could be probed in experiments in order to understand the intermediate phase. 

\section*{Model}
\begin{figure}[hbtp]
\centering
\includegraphics[width=3.3in]{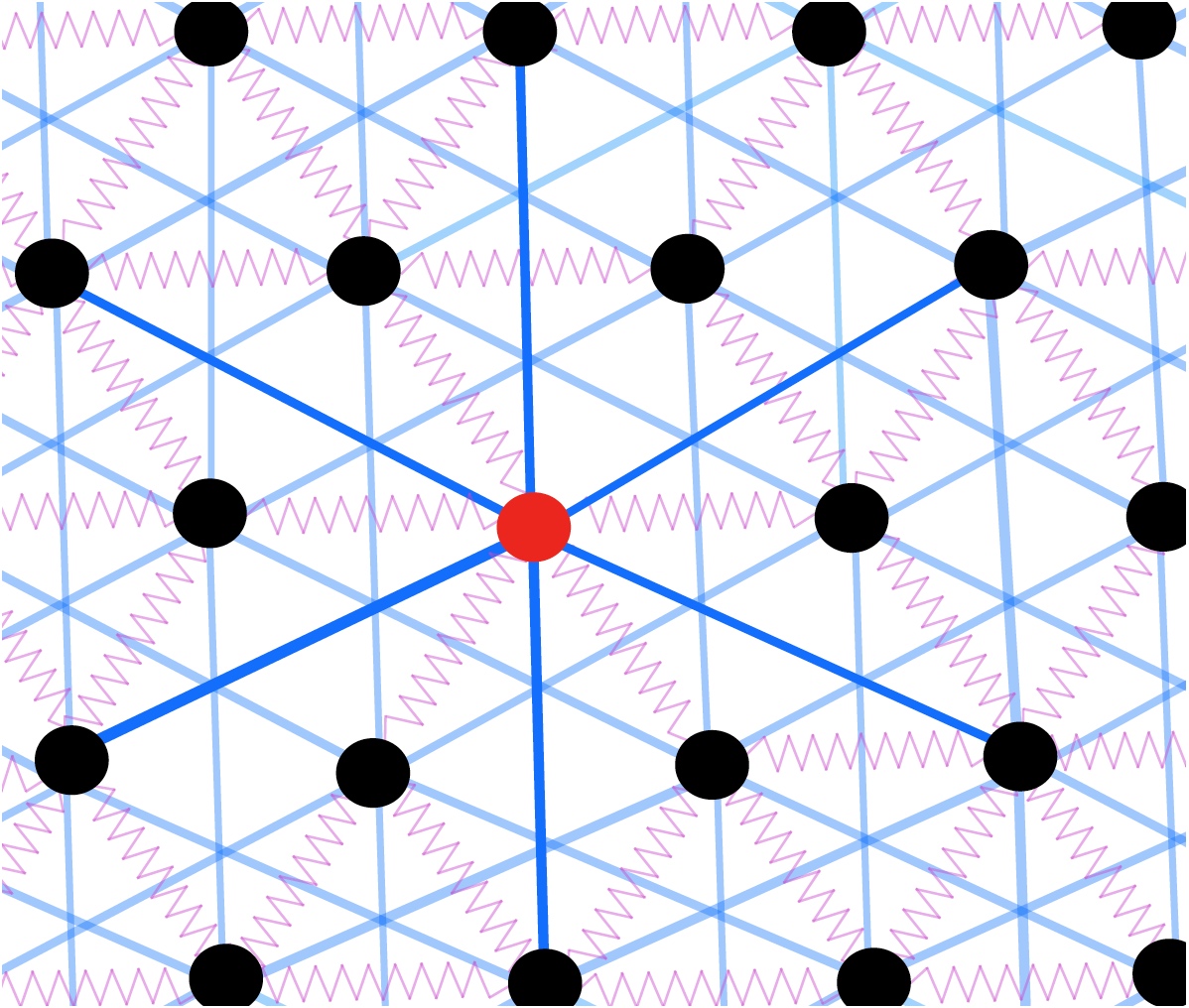}
\caption{Sketch of the model. The black circles represent the particles, the purple springs represent the strong interaction between nearest neighbors while the blue lines represent the VdW interactions between next nearest neighbors. We color one of the particles in red with bold blue lines for illustration purposes. } \label{fig:sketch}
\end{figure}

We consider a two-dimensional triangular lattice of $N$ particles with periodic boundary conditions~\cite{Yan14,Yan15,Yan18}, where a small regular deformation of the lattice is imposed to avoid non-generic singular modes. 
We model  all radial and angular constraints of covalent interactions by $N_s=N n$ linear springs of stiffness $k$, connecting the nearest neighbors on a triangular lattice, as shown in Fig.~\ref{fig:sketch}. 
We incorporate quenched disorder of glassy energy landscape by rest length mismatches of springs to the lattice bond lengths: $l_{\gamma}=l_{\gamma,0}+\epsilon_{\gamma}$ for spring $\gamma$. Mismatches $\lbrace\epsilon_{\gamma} \rbrace$ are i.i.d. random Gaussian variables with mean zero and variance $\epsilon^2$. 
By setting $k_B=1$, $k\epsilon^2=1$ defines the unit of, both, energy and temperature. 
Furthermore, we include also the weak VdW interactions by adding weak fixed springs of stiffness $k_w \ll k$ connecting any particle to all its six next-nearest neighbors as shown in Fig. \ref{fig:sketch}. The mean-field effect of these weak long-range nonspecific interactions can be captured by a control parameter $\alpha=3k_w/k \ll 1$~\cite{Yan13,Yan15}.

In the model, the microscopic configuration depends on how the network is connected or which lattice bonds are occupied by strong springs, denoted by $\Gamma \equiv \lbrace \gamma \leftrightarrow (i,j) \rbrace$, for particle $i$ and $j$ connected by spring $\gamma$. 
Given $\Gamma$, when particles deviate from the mechanical equilibrium by $|\delta{\bf R}\rangle$, the elastic energy potential to the linear order is,
\begin{equation}
V(\Gamma, |{\bf R}\rangle)= H(\Gamma)+\frac{1}{2}\langle\delta {\bf R}|{\mathcal M}(\Gamma)|\delta{\bf R}\rangle,
\end{equation}
where $H$ is the energy of the inherent structure of configuration $\Gamma$ and the second term corresponds to the vibration from equilibrium with $\mathcal{M}$ being the Hessian matrix of energy $H$. 
We thus perform a Metropolis Monte Carlo simulation~\cite{Newman99} to sample the configurations according to their Boltzmann weight $e^{-F(\Gamma)/T}$ with free energy,
\begin{equation}
F(\Gamma) = H(\Gamma)-TS_{\rm vib}(\Gamma),
\end{equation}
with the volume of thermal vibrations counted in the vibrational entropy, 
\begin{equation}
S_{\rm vib}(\Gamma)= n_cN\ln T-\frac{1}{2}\ln{\rm det}{\mathcal M}=-\sum_{\omega}\ln\omega(\Gamma)+c,
\label{eq:svib}
\end{equation}
where $\omega^2$ are the eigenvalues of the Hessian matrix. 

Without loss of generality, we assume the independence of mismatch $\epsilon_\gamma$ on the particle distances of the distorted lattice $r_{ij}$ so that the Hessian matrix becomes only a function of the occupation $\lbrace\sigma\rbrace$, where $\sigma_{ij}=1$ if particle $i$ and $j$ are connected by a spring and $\sigma_{ij}=0$ otherwise. The stress energy of the network at mechanical equilibrium can thus be computed by,
\begin{equation}
H(\Gamma) = \frac{1}{2}\langle\epsilon|{\mathcal K}-{\mathcal K}{\mathcal S}{\mathcal M}^{-1}{\mathcal S}^T{\mathcal K}|\epsilon\rangle\; ,
\label{eq:energy}
\end{equation}
where ${\mathcal K}$ is the diagonal spring stiffness matrix and ${\mathcal S}$ is the structural matrix, both depending only on occupation $\lbrace\sigma\rbrace$. The detailed derivations and expressions of these matrices and the numerical implementation are documented in the Supplementary Material Notes 1 and 2. 

\begin{figure}[hbtp]
\centering
\subfigure[]{
 \includegraphics[width=3.2in]{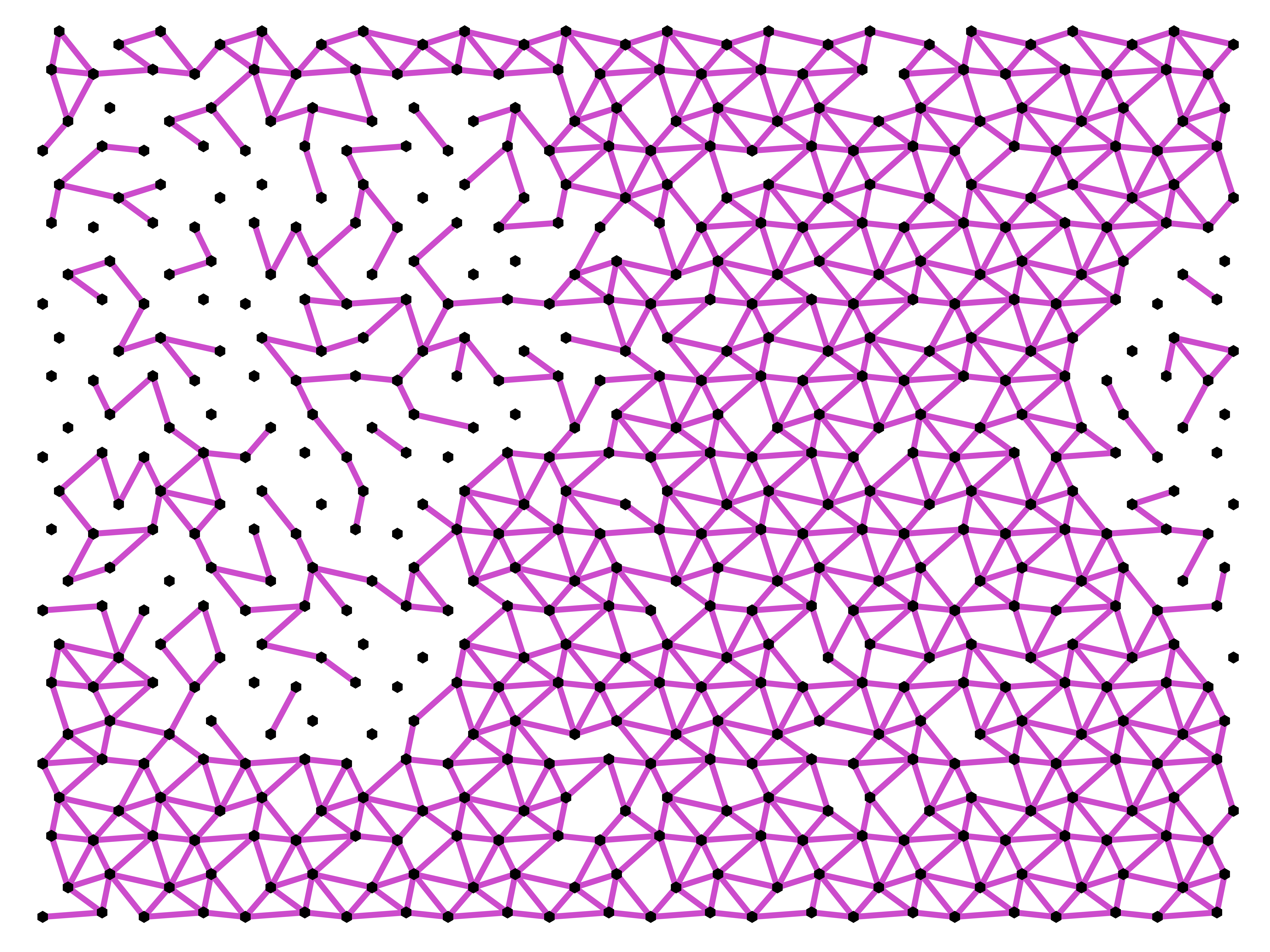}
}
\subfigure[]{
 \includegraphics[width=3.2in]{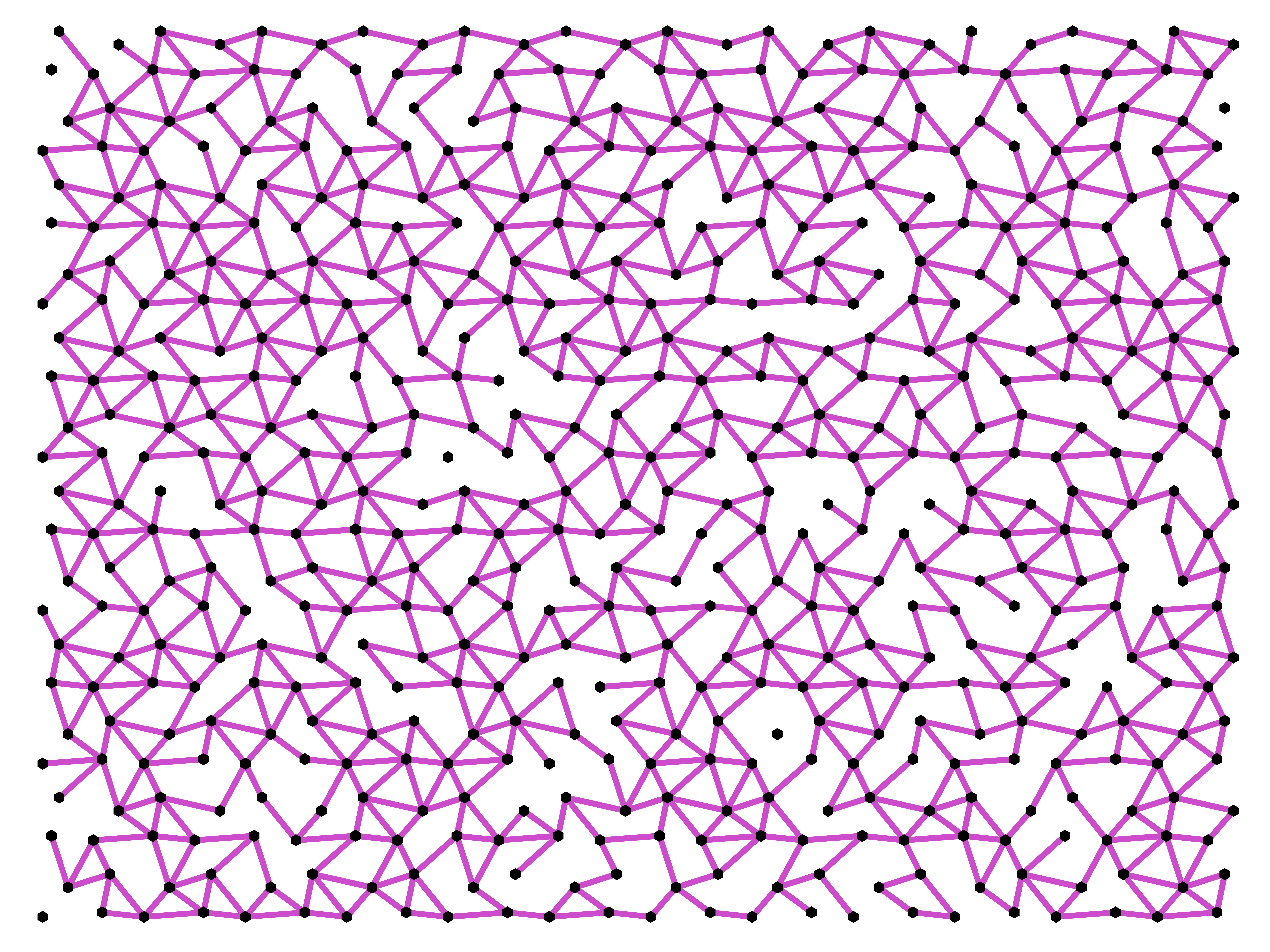}
}
\caption{Snapshot of the system above (\textbf{a}) and below (\textbf{b}) the critical temperature for a system of 576 particles with the constraint number $n=2.06>n_c$. The purple lines represent the springs. \textbf{a}) Heterogeneuous structure: At high temperature $T=10$, the entropy dominates over the internal stress energy, in particular, the vibrational entropy maximizes by phase separating into rigid and floppy regions.  \textbf{b}) Homogeneous structure: At low temperature $T=0.001$ the energy of the inherent structures dominates over the internal energy, this energy minimizes by a homogeneous distribution of constraints.} \label{fig:snapshot}
\end{figure}

\section*{Results}
\subsection*{Network structures}
As proven in Ref.~\cite{Yan18} and directly inferred by Eq.~\ref{eq:svib}, vibrational entropy is large for floppy modes with a vanishing $\omega$. When the total number of constraints is fixed, the total entropy can gain from  additional floppy modes in phase separation of a very rigid subnetwork where the springs cluster and a very floppy counterpart where floppy modes cluster.  
This phase separation is shown in the snapshot of a system of 576 particles at high temperature in the left panel of Fig.~\ref{fig:snapshot}. 
On the contrary, networks with constraints homogeneously distributed store lower elastic energy than other configurations given the number of springs, as shown in Ref.~\cite{Yan14}. 
At low temperature, when elastic energy dominates, homogeneous microscopic structures with no rigid-floppy phase separation will be sampled, as shown in the right panel of Fig.~\ref{fig:snapshot}. 
In the following, we will show that these heterogeneous and homogeneous structures correspond to two distinct thermodynamic liquid phases that are separated by a first-order liquid-liquid transition at a critical temperature $T_{LLT}$. 
We will further argue that depending on the relation between $T_{LLT}$ and the glass transition temperature $T_g$, the liquid can be frozen into different thermodynamic phases, which could be the origin of the singular intermediate phase in network glass. 

\subsection*{Thermodynamics}  
\begin{figure*}[hbtp]
\centering
\subfigure[]{%
 \includegraphics[width=3.2in]{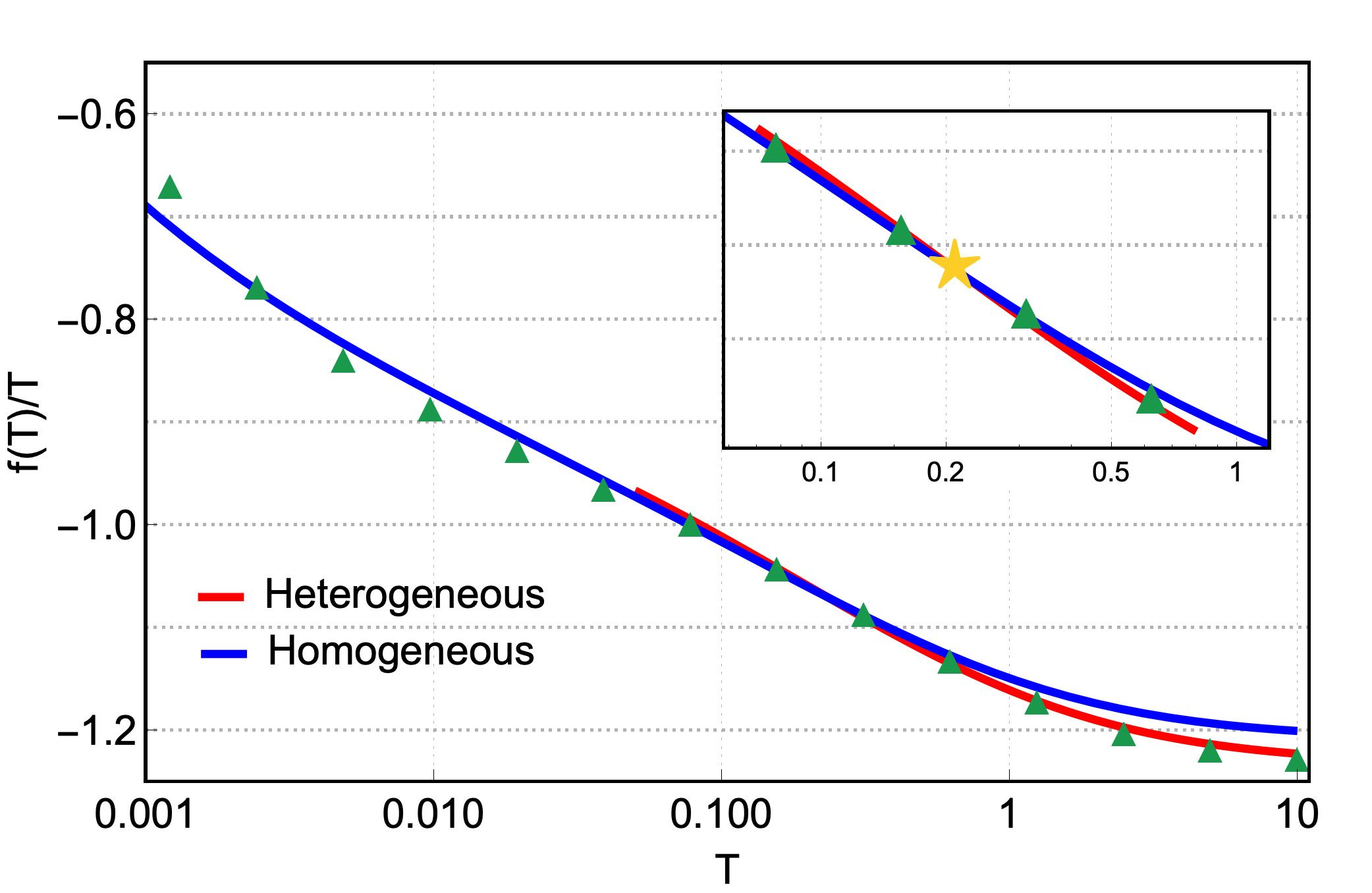}}
\subfigure[]{%
 \includegraphics[width=3.2in]{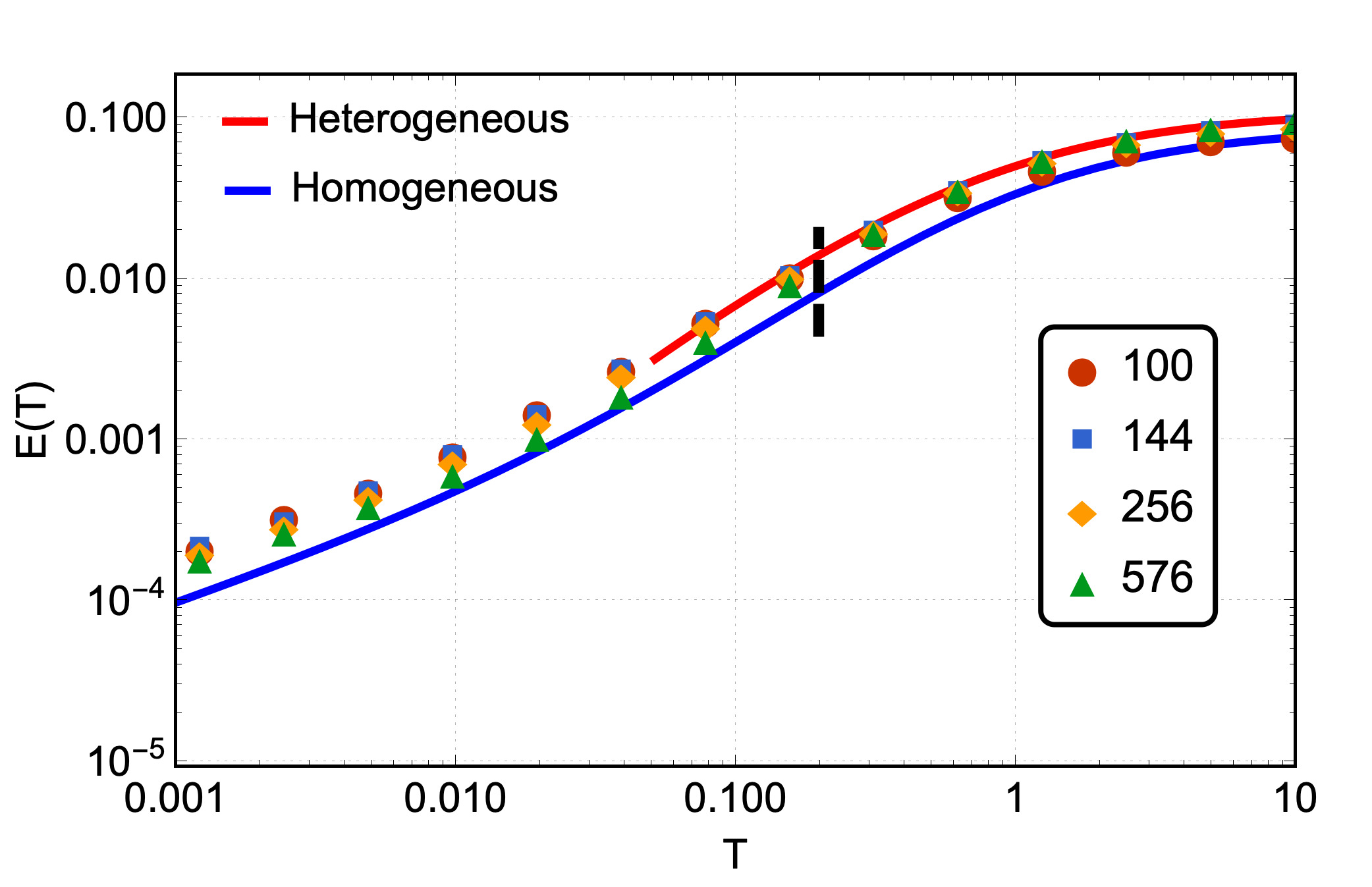}}
\\
\subfigure[]{%
\includegraphics[width=3.2in]{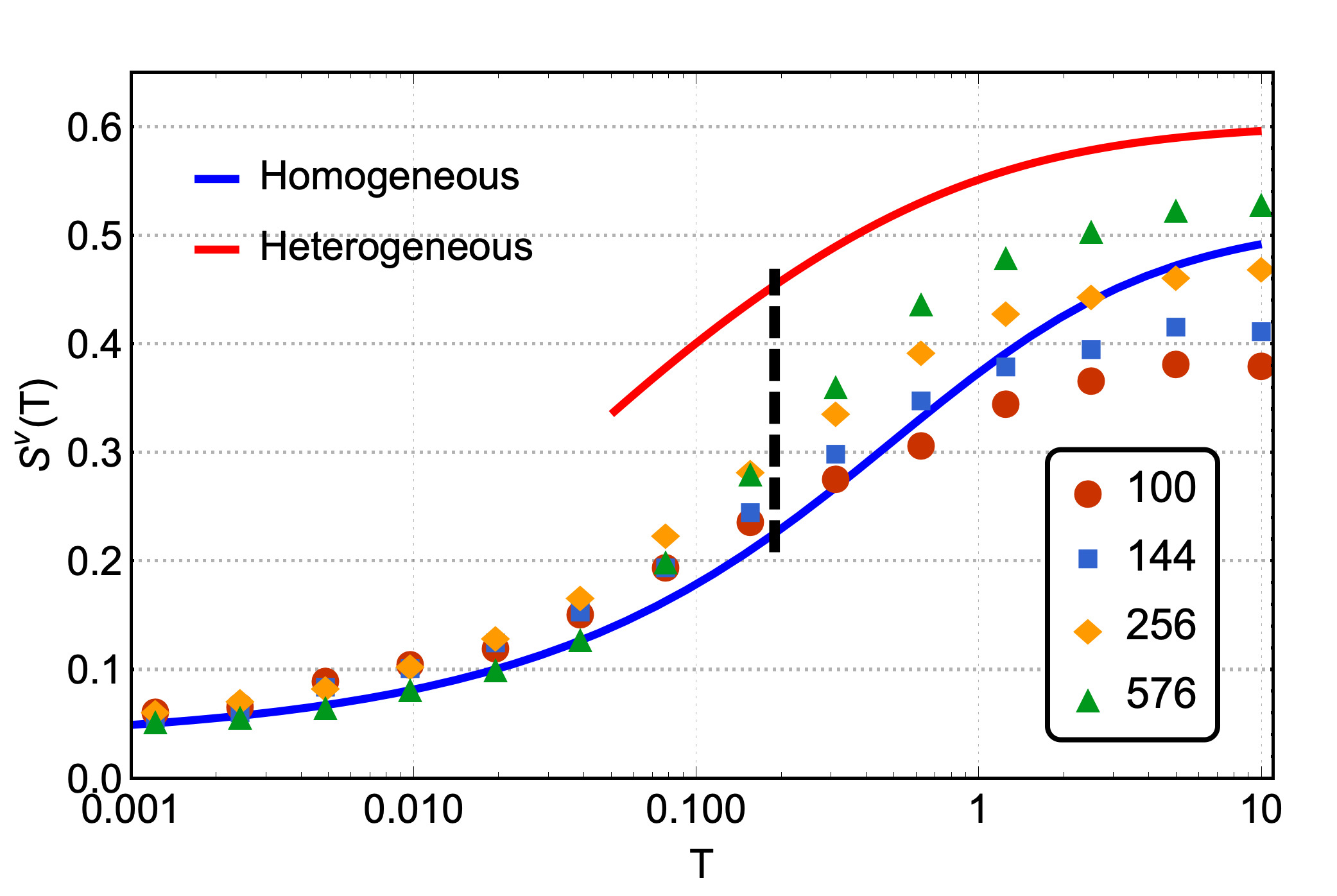}}
\subfigure[]{%
\includegraphics[width=3.2in]{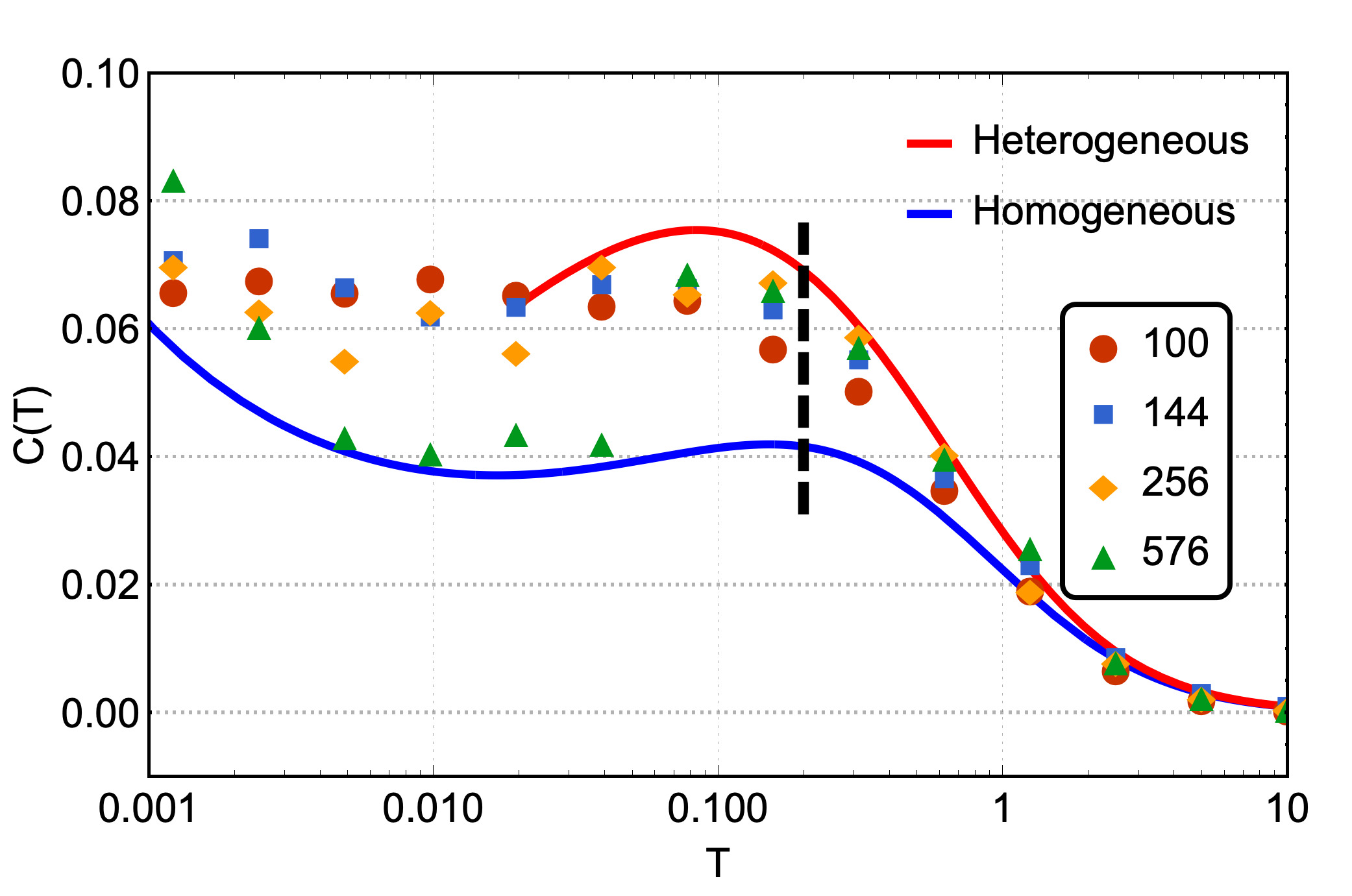}}
\caption{Thermodynamics of the network model near the rigidity transition $n=2.06$. The thermodynamics is characterized by the basic thermodynamic quantities, including free energy, internal energy, entropy, and specific heat, shown versus temperature in markers for simulation results and in curves for analytical predictions. 
The simulations are done for different system sizes $N$ (see legends) and the analytical predictions are obtained in the thermodynamic limit $N\to\infty$ for homogeneous networks in blue and heterogeneous networks in red (see Supplementary Material Notes 3$\&$4 for detailed descriptions). 
\textbf{a)} The numerical results of free energy follow the prediction of a homogeneous network at low temperatures until the homogeneous-heterogeneous first order phase transition around $T \approx 0.2$ and then cross over to the prediction of a heterogeneous network. The yellow star in the inset marks this crossover.
\textbf{b)} Data points follow the homogeneous and heterogeneous predictions in the same low and high temperature ranges corresponding to the free energy, while separated by a discrete transition that the numeric result is converging to in the thermodynamic limit. 
\textbf{c)} Similarly, the vibrational entropy results also converge to a discrete jump at the crossover of free energy. 
\textbf{d)} At the phase transition, the specific heat is also characterized by has a jump, seen in the largest system size. 
} \label{fig:thermodynamics}
\end{figure*}

The numerical results of thermodynamics of the model are shown in Fig.~\ref{fig:thermodynamics} together with the theoretical predictions of both heterogeneous and homogeneous phases.  
In the upper left panel of Fig.~\ref{fig:thermodynamics}, for a given connectivity $n=2.06$, we find the total free energy of the networks equilibrated at given temperature $T$ can be perfectly fitted by the theoretical predictions of heterogeneous networks at high temperature end (in red) and of homogeneous networks at low temperature end (in blue). 
Moreover, the numerics and the free energy prediction of heterogeneous networks are consistently lower than the prediction of the homogeneous phase when the temperature is higher than certain transition temperature $T_{LLT}\approx0.2$. 
At the free energy crossover $T_{LLT}$, marked in Fig.~\ref{fig:thermodynamics}(b)(c)(d), we are also observing the convergence to discontinuous jumps at the transition in the thermodynamic limit $N\to\infty$ from a higher value in high temperature heterogeneous phase to a lower value in the low temperature homogeneous phase in stress energy, vibrational entropy, and the specific heat. 
This result demonstrates that the heterogeneous and homogeneous structures are thermodynamical phases, separated by a first-order phase transition where, both, energy and entropy are discontinuous. 

In Fig.~\ref{fig:thermodynamics}, the data points of energy $E=\overline{H}^T$ in the upper right panel of  Fig.~\ref{fig:thermodynamics} and vibrational entropy $S_{v}=\overline{S_{\rm vib}}^T$ in the lower left panel of Fig.~\ref{fig:thermodynamics} are averages of Eqs. (\ref{eq:svib}) and (\ref{eq:energy}) over the Monte Carlo courses at given temperature $T$. 
The specific heat $C$ in the lower right panel of Fig.~\ref{fig:thermodynamics} is obtained from the mean energy fluctuation over the Monte Carlo courses normalized by temperature squared, $C=(\overline{H^2}^T-E^2)/T^2$. 
Finally, the main numerical result of free energy $F$ in the upper left panel of Fig.~\ref{fig:thermodynamics} combines both direct measurement of energy $E$ and the inferred total entropy $S=S_{v}+S_{c}$ by integrating over the specific heat $C$,
\begin{equation}
S(T)=S(\infty)-\int_T^\infty\frac{C(T)}{T}dT,
\end{equation}
as $F=E-TS$. 
The theory derivations and the way we consistently fit parameters are fully documented in the Supplementary Material Notes 3-5 or see Ref.~\cite{Yan15} for homogeneous networks and Ref.~\cite{Yan18} for heterogeneous networks. 

\begin{figure}[hbtp]
\centering
\subfigure[]{
\label{fig:Sk1}
\includegraphics[width=3.1in]{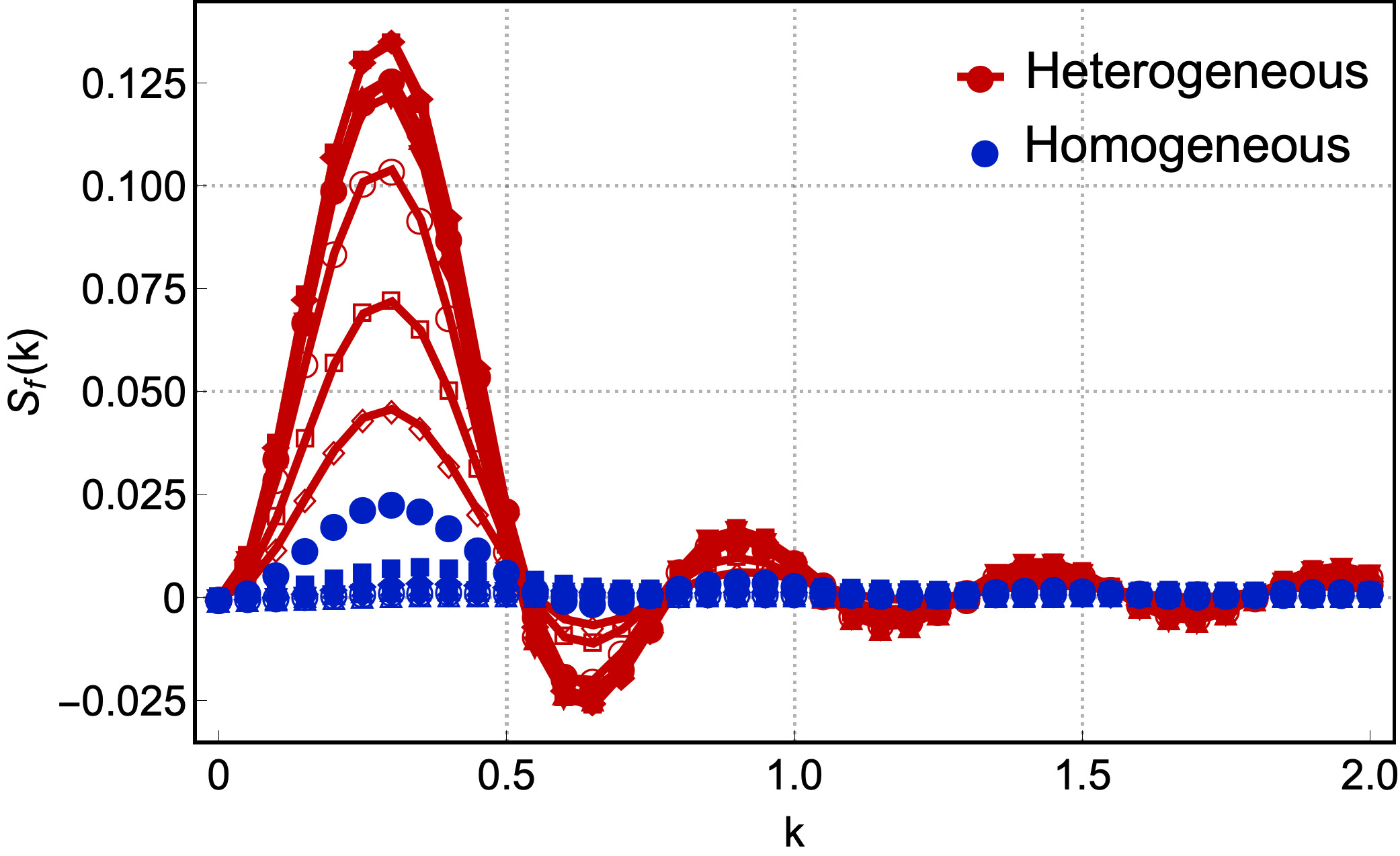}}
\subfigure[]{
\label{fig:Sk2}
\includegraphics[width=2.9in]{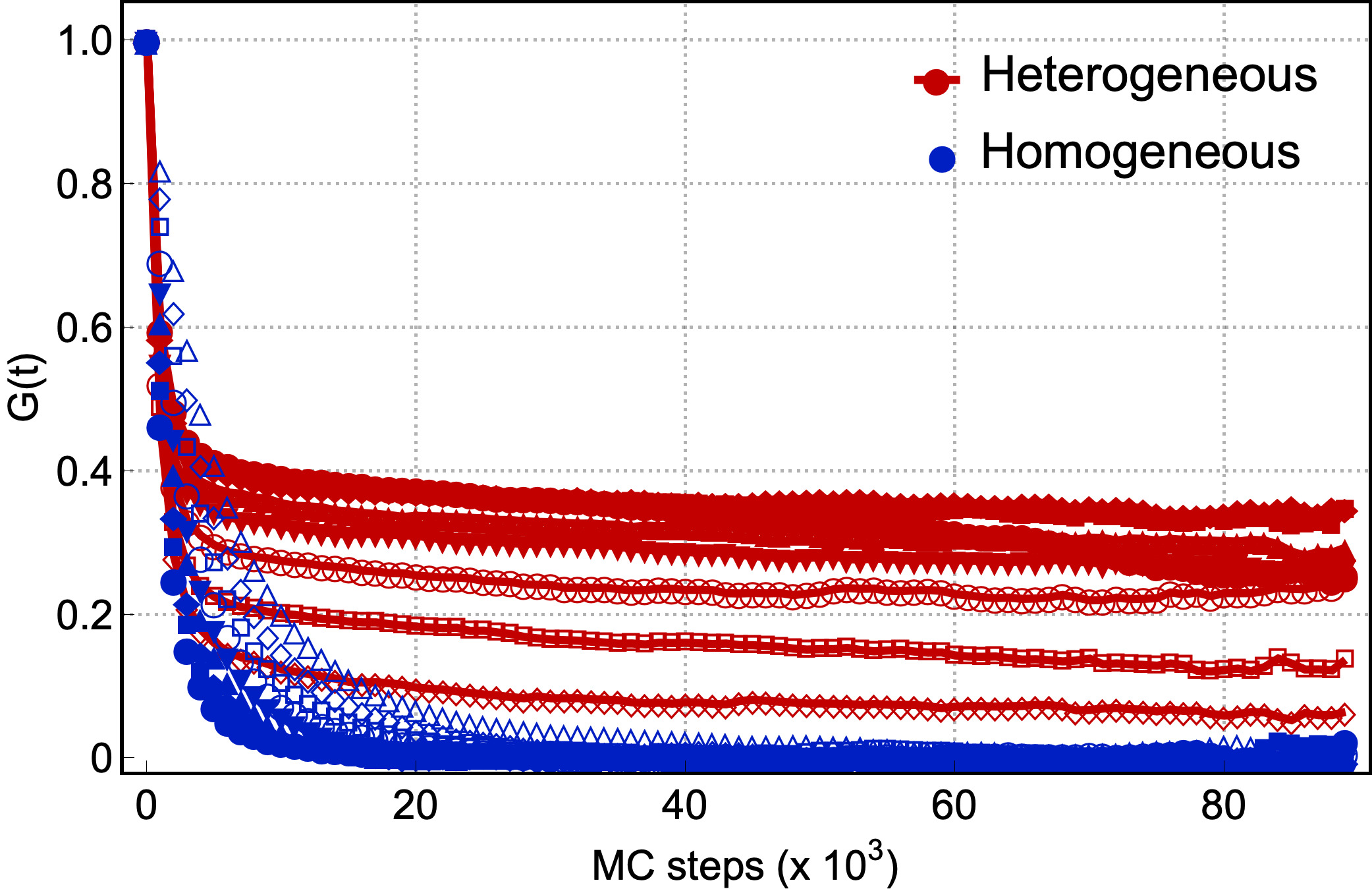}}
\caption{{\bf a}) Spatial correlation function and {\bf b}) time correlation function for $N=576$. Red curves correspond to temperatures $T(m)=10/2^m$ with $m=\lbrace 0,...,5 \rbrace$ and blue curves correspond to temperatures $T(m)$ with $m=\lbrace 6, ..., 16 \rbrace$. The blue open symbols correspond to temperatures $T \lesssim \alpha $, i.e. of the order of the weak interactions. The wavenumber $k$ has been averaged over three different directions. The time correlation function was computed using samples taken every $10^3$ Monte Carlo steps. } \label{fig:Sk}
\end{figure}

\subsection*{Spatial and temporal correlations} 
We have shown the existence of two distinct thermodynamic liquid phases of networks with the basic thermodynamic quantities. Among these quantities, the specific heat is a good experimental indicator to detect the two liquid phases and the transition: one can look for a lambda divergence or a peak in specific heat above glass transition $T_g$, as found in certain strong-type glass-forming liquids and water~\cite{Angell08, Wei11, Angell11, Wei13}. 
Here we present also the spatial and temporal correlation profiles of the two phases that could be directly measured in experiment to probe the transition. 
The spatial and temporal correlations are investigated by the structure factor and the time autocorrelation function as shown in Fig.~\ref{fig:Sk}. They are defined by the occupation $\lbrace \sigma\rbrace$ as,
\begin{equation}
S_f(k) = \frac{1}{3N(3N-1)}\sum_{ij\neq kl}(\sigma_{ij}-\bar\sigma)(\sigma_{kl}-\bar\sigma)e^{i k r_{ij,kl}},
\end{equation}
\begin{equation}
G(t) = \frac{1}{T_{tot}-t}\sum_\tau\frac{1}{3N}\sum_{ij}[\sigma_{ij}(\tau)-\bar\sigma][\sigma_{ij}(\tau+t)-\bar\sigma],
\end{equation}
where $\bar\sigma={n}/{3}$. 

For temperatures higher than the transition temperature $T_{LLT}\approx0.2$, we observe a plateau to finite correlation in the time range scanned in simulation and a strong signal in structure factor averaged over that time scale, which reflects the heterogeneous phase as in the snapshot in the left panel of Fig.~\ref{fig:snapshot}. 
On the contrary, for temperatures lower than $T_{LLT}$, we observe normal homogeneous liquid, where the correlation quickly relaxes to zero with no special structure in wave vector space after averaged over time, as in the snapshot in the right panel of Fig.~\ref{fig:snapshot}. 
In the left panel of Fig.~\ref{fig:Sk}, we also notice that the systems in the heterogeneous phase yield two relaxation times: the system first relaxes to a plateau rapidly, yet in this plateau, the system is also relaxing but with a much larger characteristic time. It implies that the rigid and floppy clusters are not held fixed in a given position and the structural features in $S_f(k)$ will also vanish when averaged at the time longer than the second relaxation as in liquids. 
These features of spatial and temporal correlations emerging at an intermediate time scale should be looked for in distinguishing the two liquid phases and detecting the transition.
 
\begin{figure}[hbtp]
\centering
\subfigure[]{
 \includegraphics[width=2.8in]{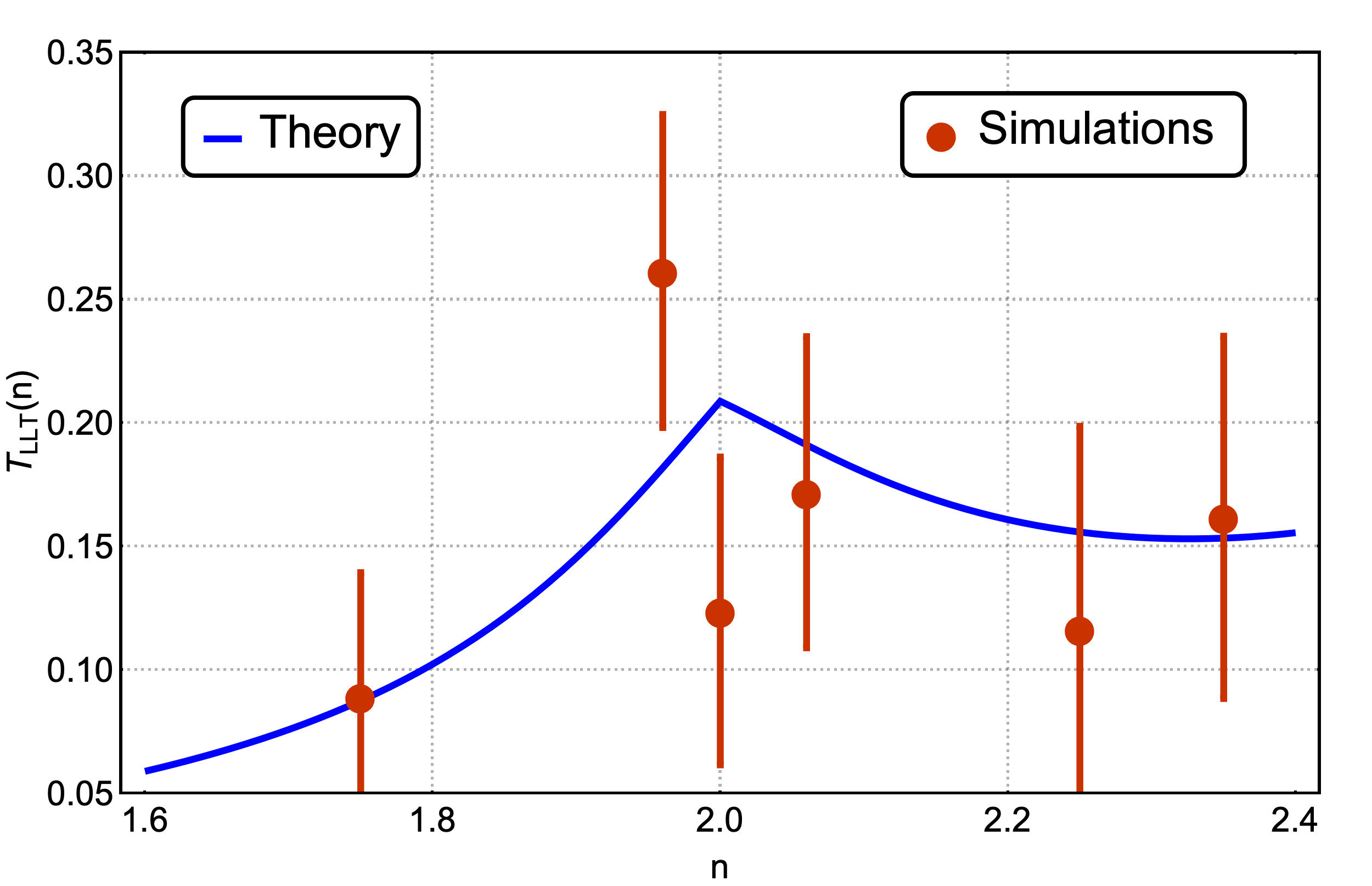}}
\subfigure[]{
 \includegraphics[width=3.2in]{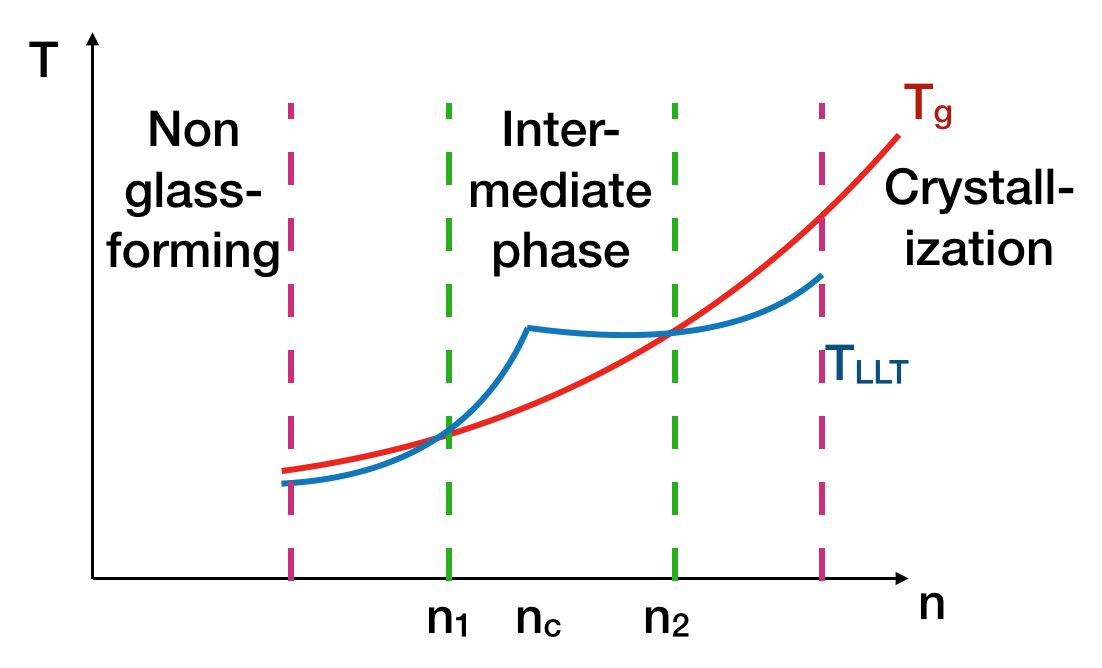}}
\caption{ {\bf a}) Liquid-liquid transition temperature vs number of constraints, predicted by theory (blue line) and  measured numerically for the network model for $N=256$ (data points). Details of the theory and numerical extraction of $T_{LLT}$ are documented in Supplementary Material Note 6. {\bf b}) {Illustration of different dependence of $T_{LLT}$ and glass transition $T_g$ on the number of constraints $n$. When $n$ is close to $n_c$ where $T_{LLT}$ becomes greater than $T_g$, liquid is frozen in a homogeneous intermediate phase at $T_g$.} } \label{fig:TLLT}
\end{figure}

\subsection*{Dependence on network topology}
Finally, we derive the liquid-liquid transition temperature $T_{LLT}$ for varying constraint number $n$ but close to $n_c$ where a heterogeneous phase exists at high temperature, shown in Fig.~\ref{fig:TLLT}(a)~
\footnote{{Notice that the temperature predicted here is in the unit of covalent bond bending and stretching energy. For instance, in the case of silica with an average constraint number $n_{{\rm SiO}_2}=3.67$, the liquid-liquid phase transition temperature has been experimentally reported at $T_{LLT}\approx1820K$ ~\cite{Brueckner70,Horbach99} and glass transition at $T_g\approx1425K$. The bond energy is estimated at $621.7 kJ/mol$~\cite{Jutzi07}, and the bond bending/stretching energy can be estimated by Lindemann's criterion with $k\epsilon^2=2\times 0.3^2\times621.7\approx112kJ/mol$.  We then have $T_{LLT}\approx0.14$ and $T_g\approx0.11$ in the unit of $k\epsilon^2$ for silica.}}.
Unlike the glass transition temperature $T_g$, which increases monotonically with $n$, $T_{LLT}$ varies non-monotonically and is maximal at $n=n_c$, which is also consistently shown by numerical results of the model as data points in Fig.~\ref{fig:TLLT}(a).  
This result implies that for certain range of parameters, the networks undergo LLT to a thermodynamic homogeneous phase before they are dynamically trapped in glass, when $T_{LLT}(n)>T_g(n)$ or $n_f<n<n_r$, which is likely to occur in the vicinity of the rigidity threshold $n_c$ due to the different dependences of $T_{LLT}$ and $T_g$ on $n$. 
The liquids frozen in homogeneous networks become glass in the IP, 
while the network glass out of the IP is then frozen in the heterogeneous network structures as the glass transition happens first under cooling, as illustrated in Fig.~\ref{fig:TLLT}(b). 
The transitions to the IP are thus transitions between different frozen thermodynamic liquid phases in this picture.  

\section*{Discussion}
Relying on how and where $T_g$ and $T_{LLT}$ intersect with each other, this new picture of the intermediate phase is potent to explain some of the material features in the experiments.
First, depending on the relative strength of the Van der Waals forces, the constraint numbers where $T_g$ and $T_{LLT}$ intersect vary, which leads to different widths and locations of the intermediate phase when changing the chemical compositions~\cite{Boolchand01, Yan18}. 
Second, as the dynamics have shown to be much less fragile in a liquid with homogeneous structures,  the liquid-liquid transition from the high-temperature heterogeneous to low-temperature homogeneous phase implies the dynamics of a liquid in the intermediate phase potentially undergoes a fragile to strong transition under cooling as observed in water~\cite{Angell11} and {\it in-silico} silica~\cite{Horbach99,Sastry03}.
Finally, as a byproduct of our theory, the disappearance of heterogeneous phases at very high and very low $n$ may explain the transitions beyond the intermediate phase far from the rigidity threshold~\cite{Bhosle12}, as depicted in Fig.~\ref{fig:TLLT}(b). 

To test this picture of the intermediate phase experimentally, 
one could look for direct signals of the liquid-liquid transition, including a lambda peak in the specific heat and loss of structural features  from the scattering experiments under cooling. 
The direct evidence should be most likely to be found in compounds close to the boundaries of the intermediate phase, where the liquid-liquid transition temperature is comparable to the glass transition temperature. 
As the glass transition reflects the dynamic aspect while the liquid-liquid transition reflects the thermodynamic aspect of the material, one could also tune one of the transitions by increasing the cooling rate, or adding a small amount of impurities, or exerting a certain amount of pressure to check if the range of the intermediate phase can be perturbed in a predictable way. 

\section*{Conclusions}
In this paper, we have shown with an elastic network model that the microscopic structure of a network glass undergoes a liquid-liquid transition from an entropy-dominated heterogeneous phase to an energy-dominated homogeneous phase under cooling. 
At this first-order transition, the specific heat diverges, structural features disappear, and relaxation plateau vanishes. 
The transition temperature scales as the average frustration energy stored in covalent bonds and varies non-monotonically on the network connectivity.
As the glass transition temperature scales positively with the connectivity, the two transition temperatures could cross at two constraint numbers. 
Inside the two constraint numbers, we would observe the liquid-liquid transition first under cooling and obtain homogeneous network glass at glass transition as in the intermediate phase. 

\section*{Author Contributions}
JQTM and LY contributed conception and design of the study; JQTM implemented simulation; JQTM and LY developed the theory and performed the statistical analysis; JQTM and LY drafted the manuscript. Both authors contributed to manuscript revision, read and approved the submitted version.

\section*{Funding}
JQTM acknowledges a doctoral fellowship from CONACyT as well as support from DGAPA-UNAM project IN102717. LY is supported by the Gordon and Betty Moore Foundation under Grant No. GBMF2919. This research was supported in part by the National Science Foundation under Grant No. NSF PHY-1748958.

\section*{Acknowledgments}
We gratefully thank G.G. Naumis and M. Wyart for discussions. We thank the editors for invitation to contribute to the topic. 

\section*{Supplemental Data}
 \href{https://www.frontiersin.org/articles/10.3389/fmats.2019.00196/full#supplementary-material}{Supplementary Material} containing detailed theory derivations and two figures.

\bibliography{mybib}

\end{document}